# Fabrication of high electron mobility and high photoluminescence quantum yield nanoscrolled monolayer MoS$_2$

C. Abinash Bhuyan,[1,*] Kishore K. Madapu,[2,*] K. Prabakar[1], K. Ganesan[1], A. Pandian[3], and Sandip Dhara[1,*]

[1]Surface and Sensors Studies Division, Indira Gandhi Centre for Atomic Research, A CI of Homi Bhabha National Institute, Kalpakkam-603102, India

[2]Surface and Sensors Studies Division, Materials Science Group, Indira Gandhi Centre for Atomic Research, Kalpakkam-603102, India

[3]Defects and Damage Studies Division, Materials Science Group, Indira Gandhi Centre for Atomic Research, Kalpakkam-603102, India

[*]Corresponding authors e-mail id: bhuyan.2d@gmail.com, madupu@igcar.gov.in, dhara@igcar.gov.in

*Abstract*

We fabricated the 1D nanoscrolled monolayer MoS$_2$ (1L-MoS$_2$) with superior characteristics from 1L-MoS$_2$ film in a facile route, using a suitable organic solvent with optimum surface tension, evaporation rate and dielectric constant, which facilitates the controlled scroll formation. These nanoscrolls behave as multilayers in morphology and monolayer electronically. The nanoscrolls exhibited a direct optical gap with enhanced photoluminescence quantum yield stemming from the weak interlayer coupling among constituent layers and were corroborated by low-frequency Raman measurements and Kelvin probe force microscopy measurements. Furthermore, enhanced photoluminescence emission after annealing uncovers the thermal stability of nanoscrolls. In addition, conducting atomic force microscopy results exhibit a significantly higher photocurrent in the nanoscrolled 1L-MoS$_2$ compared to the 1L-MoS$_2$. We also realized significantly improved field effect transistor device parameters in nanoscrolled 1L-MoS$_2$ devices. In nanoscrolled devices, we report the highest mobility value of 2400 cm$^2$V$^{-1}$s$^{-1}$ reported in any form of 1L-MoS$_2$.

# 1. Introduction

Ultra-thin monolayer $MoS_2$ (1L-$MoS_2$) attracted a great deal of attention among the research community because of its interesting electronic and optical properties, with the observation of direct band gap and enhanced photoluminescence (PL) quantum yield (QY).[1] The PL QY of 1L-$MoS_2$ can be enhanced by doping,[2] surface treatment,[3] and increasing the heat dissipation area.[4] Recently, it has been reported that PL QY of 1L-$MoS_2$ can be enhanced with an increase in the flake area owing to the availability of a large heat-dissipating area.[4] In particular, large-area 1L-$MoS_2$ provides one-order higher PL QY than mechanically exfoliated 1L-$MoS_2$ and two-order higher than small-sized as-grown chemical vapour deposition (CVD) grown flakes.[4] At the same time, devices fabricated with large channel dimensions may not meet the required demand to fabricate miniaturized and compact next-generation electronic devices. In other words, there is a trade-off between the availability of large heat-dissipating areas and the compactness of the devices. Therefore, it is imperative to eliminate the bottleneck by designing a compact structure that retains the characteristics of large-area 1L-$MoS_2$. It is worth mentioning that the optical absorption coefficient in the visible wavelength of the electromagnetic spectrum of 1L-$MoS_2$ is significantly high compared to that for conventional semiconductors.[4] However, the overall light absorption is limited by the atomic scale thickness of monolayer $MoS_2$. Thus, the structure containing multiple layers possessing the monolayer behaviour will be necessary for a significant amount of absorbed light in the visible range.

Recently, the one-dimensional (1D) nanostructures designed from 2D monolayers have opened new avenues for tuning their structural, electronic, and optical properties.[5,6]

According to the theoretical calculations, the nanoscrolls have a unique topology, which makes them useful in flexible electronics, energy storage, and optical resonator applications.[7] The scrolled 1L-MoS$_2$ gained substantial attention because it exhibited higher light-matter interactions, which resulted in higher photo-responsivity than the planar 1L-MoS$_2$.[8] In scrolled 1L-MoS$_2$, the transportation of the carriers occurs in their constituent layers, which is quite different from their respective multilayers.[9] For the first time, amorphous MoS$_2$ nanoscrolls were prepared by the Ar plasma-assisted method.[10] The PL properties of nanoscrolled MoS$_2$ vary significantly depending on the fabrication process.[8,9,11-13] Table S1 summarizes the optical properties of reported nanoscrolled MoS$_2$ samples. So far, the nanoscrolls were mostly fabricated by the self-rolling method. The self-rolling was initiated by the evaporation of volatile organic solvent over the monolayer. Usually, isopropyl alcohol (IPA)[12] and acetone[11] were used as organic solvents because of their low surface tension and high evaporation rates. However, the PL emission studies were absent in nanoscrolls fabricated using IPA.[8,12] In contrast, the self-rolled nanoscrolls fabricated with the aid of acetone showed four times higher PL intensity.[11] Unfortunately, PL emission in these samples was quenched after annealing (100 °C for 1h). In addition, capillary force-driven self-rolls of vertical van der Waals heterostructures (SnS$_2$/WSe$_2$) were also prepared using a mixture of ethanol, water, and ammonia.[5] However, a reduction in PL emission was observed in heterostructured scrolls compared to planar structures, and it was attributed to their intimate interlayer interactions. Alternatively, the mechanical force-induced scroll fabrication has also been investigated using a mixture of ethanol and pure water.[8,9] The strain-induced MoS$_2$ nanoscrolls fabricated using ethanol solution (volume ratio of ethanol: water = 2:1) showed properties of multilayered MoS$_2$ with reduced PL QY.[9,13] Scrolls fabricated by wtheater-drag method lack PL measurements.[8] In addition, the mechanical force-induced rolls are mostly prone to structural deformity due to stacking disorder and non-uniform rolling. So far, PL emission was either not observed or quenched after annealing in earlier reported nanoscrolled MoS$_2$.[8] Thus, fabrication of nanoscroll MoS$_2$ retaining the monolayer properties and maintaining thermal stability also is necessary for device applications.

In this report, we developed and demonstrated the effective and facile route to fabricate the thermally stable nanoscrolled 1L-MoS$_2$. The self-rolled nanoscrolls were fabricated by using absolute ethanol. The scrolled 1L-MoS$_2$ exhibits a high quantum yield of 2.95% and up to five times more PL intensity than remnant 1L-MoS$_2$. The enhancement in the PL emission was attributed to negligible interlayer coupling among constituent layers. Low-frequency Raman and Kelvin probe force measurements confirm negligible interlayer coupling. From the PL analysis, it was found that interlayer coupling in nanoscrolls further weakened after annealing. The annealed nanoscrolls have shown nearly six times higher PL intensity compared to its counter planar 1L-MoS$_2$ structures. The nanoscrolled 1L-MoS$_2$ showed significantly improved electronic and optoelectronic properties compared to their planar counterparts.

**Self-rolled nanoscrolled 1L-MoS$_2$ fabrication**

The nanoscrolled 1L-MoS$_2$ was fabricated from CVD-grown 1L-MoS$_2$ film. The as-grown large-area 1L-MoS$_2$ on SiO$_2$/Si substrate was taken as the starting material. A little (15-20 μl) amount of absolute ethanol (Sigma Aldrich, purity 99.9 %) was dropped on the sample surface containing 1L-MoS$_2$. The optical microscope captured the step-by-step image of an exemplary nanoscrolled 1L-MoS$_2$ at the formation stage (Figure 1). In addition, nanoscrolled 1L-MoS$_2$ fabrication steps were provided schematically in supplementary (Figure S1). Figure 1a-c illustrates the optical micrographs at each stage of the nanoscroll formation. Figure 1a depicts the ethanol drop on the planar 1L-MoS$_2$ surface. Subsequently, the size of the ethanol drop shrinks due to evaporation at room temperature (RT) (Figure 1b). The complete evaporation of ethanol generates the rolled structures in a few areas (Figure 1c). The high magnification image of one of the fabricated nanoscrolls was provided in Figure 1d. Moreover, the optical contrast between 1L-MoS$_2$ and rolled structure exhibits the successful formation of scrolled MoS$_2$ (Figure 1d). Immediately after the formation of the scroll, sample was placed under an IR lamp for 1 h to remove any ethanol residues from MoS$_2$ surface. Interestingly, scroll formation was only observed in certain regions of the sample. In particular, scroll formation was initiated at the grain boundaries region of large area MoS$_2$ film and edges of triangular flakes. Thus, one can speculate that the scroll formation might have happened because of the penetration of ethanol between the flake and substrate at the grain boundaries and edge region and subsequent evaporation. Ethanol delaminated the 1L-MoS$_2$ during the evaporation at RT. During the evaporation of ethanol, the heat might be extracted from the planar 1L-MoS$_2$ structure, which resulted in reduction of surface energy.[9] The decrease in surface energy is compensated by the change of planar MoS$_2$ structure to its scrolled form.

In general, the scroll formation process is governed by solid-liquid interfacial energy and the solvent surface tension and evaporation rate.[5] The interfacial energy is solely responsible for the intercalation of solvent molecules to delaminate the 1L-MoS$_2$ from the SiO$_2$ substrate. The surface tension and evaporation rate of the solvent are two important parameters as they decide the capillary force (flow direction of the solvent during evaporation) at the scroll formation stage.[5] In earlier reports, the mixture of ethanol and water, IPA, and acetone solvents were used for the scroll fabrication. Unfortunately, PL emission was either suppressed or reduced after annealing in these scrolls compared to the planar MoS$_2$.[11] Here, we fabricated the scrolls using absolute ethanol because of its optimum physical properties, such as surface tension and evaporation rate. Generally, the mixture of ethanol and water has a higher surface tension (~72.8 mN·m$^{-1}$ of water-air @ 20 °C) than that of pure organic solvents like ethanol (~22.1 mN·m$^{-1}$ of ethanol-air @ 20 °C), IPA (~23 mN·m$^{-1}$ of IPA-air @ 20 °C) and acetone (~25.2 mN·m$^{-1}$ of acetone-air @ 20 °C). [14] The high surface tension of the solvent is prone to bead-up and deteriorates the uniformity of the scroll.[5] Moreover, ethanol has optimum evaporation rate of 119 mg·m$^{-2}$·s$^{-1}$ and vapour pressure ~7.86 kPa@25 °C compared to those of acetone (evaporation rate 504 mg·m$^{-2}$·s$^{-1}$ and vapour pressure ~30.8 kPa@25 °C) and IPA (evaporation rate 75.6 mg·m$^{-2}$·s$^{-1}$ and vapour pressure ~5.69 kPa@25 °C).[14] The optimum evaporation rate of ethanol is reasonably advantageous as it allows sufficient time for a uniform scroll formation process.

In addition, it was also reported that the formation of scrolls and their stability is governed by the thermodynamical parameters. The formation of scrolls is allowed if the enthalpy ($H$) decreases upon scrolling. The change in enthalpy, $\Delta H = \Delta U + \Delta W$, where $\Delta U$ stands for change in internal energy and $\Delta W$ stands for work done due to rolling. Most importantly, the $\Delta W$ is dependent on the dielectric constant of the solvents involved in the scroll formation. [6] At room temperature, the dielectric constant of ethanol, IPA and acetone are 24.55, 18.3, and 20.7, respectively. The dielectric constant value of ethanol is comparatively higher than IPA and acetone. Thus, the self-rolling process in an ethanol medium is relatively slower than in IPA and acetone.[15] Therefore, absolute ethanol is advantageous over other organic solvents for controlled and uniform scroll formation.

## 2. Results and Discussions

The cross-sectional view of the scrolled 1L-MoS$_2$ is represented as a schematic in Figure 2a. In the present study, we compared the optical properties of the scroll with remnant 1L-MoS$_2$ for the fair assessment. The multiple scrolls were formed using a single drop of ethanol (Figure S2). Among them, we carried out a detailed optical analysis on a particular scroll (Figure 2b). FESEM images of other scrolled 1L-MoS$_2$ structures are shown in Fig S3. Figure 2c shows the typical FESEM image of the same scroll shown in Figure 2b, and its high magnification image is provided in Figure S4a. It reveals that the monolayer is uniformly rolled to form the tubular structure. In addition, it is apparent that the fabricated scrolls are clean and wrinkle-free. The length of the given nanoscroll is measured to be ~13.5 µm. The lateral size of the scroll across length was in the range of 220-250 nm. We have carried out an AFM study to estimate the thickness of the 1L-MoS$_2$ and the height of the nanoscroll. The thickness of 1L-MoS$_2$ is 0.85 nm (Figure S4b,c). The AFM image of the typical nanoscrolled 1L-MoS$_2$ was provided in Figure 2d. The surface topography of the nanoscroll is uniform. From the height profile, the thickness of the nanoscroll was calculated to be 160±10 nm (insets of Figure2d). We calculated the number of layers present in the scroll by assuming it as tubular form, and detailed calculation steps are provided in the supplementary material. In the given scroll, approximately 28 number of 1L-MoS$_2$ layers are estimated to be present.

Raman measurements were carried out to assess the phase and number of layers present in the planar and nanoscrolled MoS$_2$. In the back-scattering configuration, the 1L-MoS$_2$ has two prominent Raman peaks corresponding to $E^1_{2g}$ and $A_{1g}$ modes.[16,17] Raman spectra of a typical planar and nanoscrolled MoS$_2$ are provided in Figure 3a. In planar MoS$_2$, the $E^1_{2g}$ and $A_{1g}$ Raman modes are observed with ca. 383.9 cm$^{-1}$ and 403.6 cm$^{-1}$, respectively. At the same time, $E^1_{2g}$ and $A_{1g}$ Raman modes are observed at 383.5 cm$^{-1}$ and 402.6 cm$^{-1}$, respectively, for nanoscrolled 1L-MoS$_2$. A small red-shift in $E^1_{2g}$ mode is attributed to the development of tensile strain due to the curved morphology in the scroll.[18] The red-shift of $A_{1g}$ (1 cm$^{-1}$) can be understood in terms of restoring force. It is well known that out-of-plane $A_{1g}$ phonon mode frequency is strongly dictated by the restoring force, whereas in-plane $E^1_{2g}$ mode is dictated by dielectric screening.[17,19,20] In the form of a scroll, the bonding between flake and SiO$_2$ happened to be broken, leading to the disappearance of restoring force originated from the bonding between SiO$_2$ and MoS$_2$. Thus, $A_{1g}$ phonon mode red-shifted in nanoscrolled MoS$_2$ as compared to planar MoS$_2$. Raman analysis was carried out on multiple nanoscrolls and the results were summarized in Table S2. In addition, we compare the frequency difference ($\Delta$) between two prominent Raman modes ($E^1_{2g}$ and $A_{1g}$), as it reveals the number of layers present in a MoS$_2$ flake.[16] In planar MoS$_2$, $\Delta$ is 19.7 cm$^{-1}$, which confirms that planar flake is indeed monolayer.[4,16] Interestingly, the $\Delta$ value in scrolled MoS$_2$ was 19.1 cm$^{-1}$. The $\Delta$ value of nanoscrolls was similar to that for monolayer flakes, which confirmed the monolayer characteristics of the scroll.[16] As mentioned before, the scroll contains approximately 28 layers. Thus, the $\Delta$ was expected to be of the bulk MoS$_2$ (>25 cm$^{-1}$).[16] It is worth mentioning that scrolls fabricated by many other groups by using different organic solvents have shown multilayer characteristics.[8,9,11-13] In their work, $\Delta$ value was in the range of 21-24 cm$^{-1}$, which confirms that the scrolls were indeed multilayer in nature (Table S1). The increased $\Delta$ values were attributed to the stacking-induced structural changes in scrolled MoS$_2$.[9] As a result, the potential application of the monolayers cannot be fulfilled through scrolls with interacting constituent layers. In the present study, monolayer characteristics ($\Delta < 20$ cm$^{-1}$) were observed in nanoscrolled MoS$_2$. In other words, in contrast to the reported works, the fabricated nanoscrolls act as a monolayer even though they contain multiple layers. The monolayer behaviour of nanoscrolled 1L-MoS$_2$ may be attributed to the non-interacting nature of constituent layers.

A 1L-MoS$_2$, being a direct band gap semiconductor, shows intense PL emission, which makes it suitable for many optoelectronic applications.[1,2] The RT PL spectra of 1L-MoS$_2$ and nanoscrolled 1L-MoS$_2$ are shown in Figure 3b. The spectra fitted with Gaussian curves were provided in the supplementary material (Figure S5 and analysis in Table S3a). In 1L-MoS$_2$, PL spectrum was majorly contributed by A-excitons (1.84 eV) over B-excitons (1.96 eV) and trions (1.80 eV) emission. Similarly, PL spectra of nanoscrolled 1L-MoS$_2$ were contributed by A-exciton (1.83 eV), B-exciton (1.94 eV) and trions (1.79 eV) emission. Interestingly, the nanoscrolled 1L-MoS$_2$ exhibits the direct band gap features, which is a primary signature of 1L-MoS$_2$. Compared to 1L-MoS$_2$, the PL spectra of nanoscrolled 1L-MoS$_2$ were red-shifted by 10 meV. The observed negligible red-shift may be attributed to the reduction in the dielectric screening upon rolling, as it becomes free-standing after being detached from the SiO$_2$ substrate.[21] Moreover, it was also reported that the red-shift in exciton peak energy was also attributed to the introduction of bending strain.[22] Previously, blue-shift in exciton peak positions in nanoscrolled MoS$_2$ were reported, and it was attributed to the existence of solvent molecules among the interlayer gaps.[11] In the present work, the negligible shift in exciton peak positions reveals the non-existence of ethanol molecules at interlayer gaps. PL analyses on various spots of a given scroll and other scrolls were provided in the supplementary material (Figure S6 and analysis in Table S3b). Furthermore, a low emission peak at 1.64 eV was observed in the PL spectrum of scrolled 1L-MoS$_2$ (Figure 3b). The low energy peak might be attributed to interlayer coupling (*I*) peak, and it is usually observed in multilayer MoS$_2$ with weak inter-layer coupling and twisting of layers.[23,24] However, *I*–peak in PL spectra of monolayer MoS$_2$ was also reported earlier under the application of external strain.[18,20] Moreover, a small curvature or bending in the monolayer MoS$_2$ film can modulate the band gap by enhancing the interlayer coupling.[18] Therefore, the origin of *I*–peak in scrolled 1L-MoS$_2$ may be traced from the developed local bending strain. Compared to 1L-MoS$_2$, the mean and maximum enhancement in PL integrated intensity of nanoscrolled 1L-MoS$_2$ is ~ 4 and ~5.2 times, respectively (Figure 3b and Table S4). The inset of Figure 3b shows the PL imaging of scrolled region. The PL imaging was carried out using A-exciton intensity. The PL intensity was high over the scroll region compared to remnant MoS$_2$. Here, the overall enhanced PL intensity in nanoscrolls is attributed to the non-interacting nature of constituent monolayers. The constituent layers of the scrolled $MoS_2$ are quite different from the bulk $MoS_2$. In general, the bilayer $MoS_2$ (2L-$MoS_2$) shows a two-order less PL intensity compared to 1L-$MoS_2$, and the emission intensity further reduces as the number of layers increases.

PL quantum yield (QY) was measured using optical absorbance and PL, as described in our earlier report.[4] Absorbance for the nanoscrolled 1L-$MoS_2$ and 1L-$MoS_2$ films were measured to be 0.62±0.08 and 0.3±0.05, respectively, and detailed in the supplementary information (Figure S7). The improved absorbance was attributed to the enhanced light-matter interaction originating from the existence of multiple layers in the scroll. The PL QY of 1L-$MoS_2$ and nanoscrolled 1L-$MoS_2$ were calculated to be 1.3% and 2.95%, respectively. The higher QY was attributed to the enhancement of PL emission in nanoscrolled 1L-$MoS_2$ compared to the planar form.

The interlayer coupling can be analyzed using shear ($S$) and layer breathing (LB) modes of $MoS_2$.[25] Usually, shear and breathing modes are not shown in Raman spectra because of the constraints of the Rayleigh rejection filter. Low-frequency (LF) Raman measurements were carried out on fabricated nanoscrolls to investigate the interlayer coupling. The LF Raman spectra of multilayered $MoS_2$ are dominated by interlayer shear ($S$ or $E^2_{2g}$) and layer-breathing (LB or $B^2_{2g}$) mode. The $S$ mode represents out-of-phase in-plane vibrations of constituent $MoS_2$ layers, whereas the LB mode represents out-of-phase out-of-plane vibrations of the $MoS_2$ layers.[26] Thus, $S$ and LB modes are always absent in 1L-$MoS_2$ due to the non-existence of an additional layer for interactions. Figure 4 shows the typical LF Raman spectra of mono-, bi-, and nanoscrolled $MoS_2$. The bilayer spectrum was provided for the comparison. The $S$ and LB modes were absent in 1L-$MoS_2$, whereas $S$ (20.6 cm$^{-1}$) and LB (36.5 cm$^{-1}$) modes were observed in bilayer (2L) $MoS_2$. Interestingly, in nanoscrolled 1L-MoS$_2$, the existence of insignificantly minute *S*-mode and the absence of LB mode reveals the existence of weak interlayer coupling.[25] Thus, the LF Raman results reinforced the PL results that interlayer coupling is negligible in nanoscrolled 1L-MoS$_2$.

The surface potential (SP) of a MoS$_2$ film is quite sensitive to the number of layers because of the interlayer screening effect.[27] We carried out Kelvin probe force microscopy (KPFM) measurements on as-grown MoS$_2$ flake, which contains the mono-, bi- and multilayer to corroborate our assumption (Figure 5a). The presence of mono-, bi and multilayer was confirmed by the Raman and PL analysis (not shown in the figure). From the contact potential difference map, it is apparent that SP varied depending on the thickness of the flake (Figure 5b). The SP of the bilayer increased by 100 mV compared to the monolayer MoS$_2$. Thus, SP would change if there is a strongly interacting layer. Subsequently, KPFM measurements were carried out on nanoscrolled MoS$_2$ and the remnant MoS$_2$. Figure 5c shows the SP map of the scrolled and monolayer region. It was found that the SP of SiO$_2$ was higher than 1L-MoS$_2$ and nanoscrolled MoS$_2$. The SP values of SiO$_2$ and 1L-MoS$_2$ were 520 and 280 mV, respectively (Figure5d). Interestingly, negligible change in SP was observed for the monolayer and scrolled region except for the edges of scrolls. The unchanged surface potential infers a negligible change in electronic properties. Importantly, the unchanged SP value among the monolayer and scrolls concludes the negligible inter-layer coupling among the constituent layers of nanoscrolled 1L-MoS$_2$.

To investigate the thermal stability of the as-fabricated scroll, we have carried out vacuum annealing at 200 °C for 2h. We restricted the annealing temperature to 200 °C to avoid the defect introduction to the 1L-MoS$_2$ system.[4,28] The PL spectra of the annealed sample are shown in Figure 6a. The integrated PL intensity was increased by ~6.4 times compared to remnant 1L-MoS$_2$. Moreover, annealed scrolled 1L-MoS$_2$ shows ~ 5 times higher integrated PL intensity compared to as-fabricated scroll (Figure 3b). The enhancement of PL intensity further ascertains the weakening of interlayer coupling among constituent layers. To date, all the studies have failed to fabricate thermally stable nanoscrolls of 1L-MoS$_2$.[8,9,11-13] Here, we achieve the formation of thermally stable scrolls because of the selection of absolute ethanol over other solvents. In this context, as discussed earlier, the vacuum annealing of MoS$_2$ nanoscrolls at 100 °C for 1h with Ar atmosphere was previously reported.[11] The PL integrated intensity was quenched ~ 4 times with significant red-shift. In contrast, we found a negligible blue shift of 10 meV in the A-exciton peak position after annealing. Thus, our PL results of annealed 1L-MoS$_2$ nanoscrolls signify that scrolls are devoid of ethanol residues because of negligible shifts in peak positions and enhanced intensity. We also observed that the interlayer peak in the nanoscrolled 1L-MoS$_2$ disappeared after annealing. It demonstrates further weakening of the interlayer coupling among their constituent layers. The reduction in interlayer coupling after annealing may also contribute to enhanced PL intensity.

We have carried out Raman analysis on annealed MoS$_2$ scrolls to further investigate the annealing effect. The Raman peak positions of $E^1_{2g}$ Raman mode are 383.7 and 384.0 cm$^{-1}$ for monolayer and scrolled 1L-MoS$_2$, respectively. The small blue-shift in $E^1_{2g}$ mode in scrolled 1L-MoS$_2$ may be because of possible molecular rearrangement in the constituent layers with the annealing treatment.[16] Similarly, the $A_{1g}$ Raman mode in 1L-MoS$_2$ and scrolled 1L-MoS$_2$ was found at 402.8, and 404.0 cm$^{-1}$, respectively (Figure 6b). After annealing, the Δ value of scrolled 1L-MoS$_2$ was 20 cm$^{-1}$ which confirmed the retaining of monolayer nature even after annealing. Thus, the fabricated nanoscrolls were thermally stable. The $E^1_{2g}$ and $A_{1g}$ phonon modes were blue-shifted by 0.5 cm$^{-1}$ and 1.4 cm$^{-1}$, respectively, compared to the as-fabricated scroll. The observed blue-shift was attributed to the relaxation of local bending strain which was developed during the nanoscroll formation stage.[22] It is also well known that the $A_{1g}$ mode is sensitive to carrier density, and the hardening of $A_{1g}$ was reported with decreased carrier density.[17] The blue-shift of $A_{1g}$ mode indicates the reduced carrier density after annealing. The increased PL intensity after annealing may be attributed to the reduction in carrier density, which decreases the Auger recombination. The effect of annealing on $S$ and LB modes is probed by LF Raman measurements (Figure 6c). It was found that $S$ and LB modes were absent in nanoscrolls after annealing. Thus, the minor signal of $S$ mode observed in the as-fabricated scroll (Figure 4) disappeared after annealing. The absence of $S$ and LB modes in an annealed sample further confirms the negligible effect of annealing on the non-interacting nature of the constituent layers.

I-V measurements were carried out using conducting atomic force microscopy (C-AFM) under dark and illumination conditions to study the photo-response of the annealed nanoscrolled 1L-MoS$_2$. The illumination was performed using the feedback laser (wavelength ~ 650 nm) source. The detailed measurement steps are provided in the experimental section. The inset of Figure 7a represents the schematic of the C-AFM measurement set-up. I-V characteristics of the 1L-MoS$_2$ and nanoscrolled 1L-MoS$_2$ without illumination are provided in Figure 7a. The non-linear behaviour of I-V graphs represents the formation of the Schottky junction between the Pt-tip with MoS$_2$ layers. Pt and Au electrodes form Schottky and Ohmic junctions, respectively, with MoS$_2$ material.[29] Further, I-V curves also indicate the *p*-type behavior of both 1L-MoS$_2$ and nanoscrolled 1L-MoS$_2$. The origin of *p*-type behaviour stemmed from the induced doping by the Pt tip used in C-AFM measurements. In general, the dark I-V characteristics are nearly the same for both 1L-MoS$_2$ and nanoscrolled 1L-MoS$_2$. The result is obvious because the constituent layers in the nanoscrolls are not interacting with each other. Under illumination in the forward bias condition, the measured current increases approximately from - 0.5 to - 0.8 nA and - 0.5 to - 2.24 nA at -4.5 V in 1L-MoS$_2$ and nanoscrolled 1L-MoS$_2$, respectively. Here, the scrolled 1L-MoS$_2$ shows approximately six times higher photocurrent as compared to the 1L-MoS$_2$. In addition, under positive bias, the measured photocurrent in scrolled 1L-MoS$_2$ is significantly higher as compared to that of 1L-MoS$_2$. The higher net photocurrent in nanoscrolled MoS$_2$ may have originated from its higher absorption cross-section due to the availability of multiple monolayers. Each monolayer generated the photo-carriers and contributed to enhancement in the overall current. Along with point I-V measurements, we have also carried out a current map at a constant voltage of -3.5 V. Figure 7c and Figure 7d represent the current mapping of the nanoscrolls under dark and illumination, respectively. In the dark, the measured current is ~ 100 pA in the region of 1L-MoS$_2$ (lower right corner) and a relatively higher current over the scroll region (nA). These current values slightly increase under illumination. To sum up, the photocurrent in the scrolled region is higher as compared to remnant 1L-MoS$_2$. The topography of a typical scroll is provided in the inset of Figure 7b.

FET devices were fabricated on planar 1L-MoS$_2$ and nanoscrolled 1L-MoS$_2$ to evaluate the electrical properties. The detailed fabrication protocols were provided in the experimental section. After fabricating the multiple FETs, we measured both their output and transfer characteristics. In the output characteristics, the gate voltage ($V_{gs}$) was varied from 0 V to 20 V with an interval of 5 V (Figure 8a and Figure 8b). In both planar and scrolled 1L-MoS$_2$ based FET devices, $I_{ds}$ were found to raise linearly up to $V_{ds}$ of ~4 V and thereafter saturated. The observed $I_{ds}$ versus $V_{ds}$ results in both the devices are typical behaviour of FET devices. At $V_{gs}$ of 20 V, the saturation drain currents are 3 and 45 µA/µm for planar and nanoscrolled 1L-MoS$_2$, respectively. The one-order (15 times) higher drain current is attributed to the availability of multiple paths for transporting the electrons in constituent layers in scrolled 1L-MoS$_2$ compared to 1L-MoS$_2$. For similar device configurations, the saturation drain currents reported in literature are 1.35 and 0.14 µA/µm for planar 1L-MoS$_2$.[30,31] Moreover, we have also plotted the transfer characteristics of both planar and scrolled 1L-MoS$_2$ to extract the threshold voltage, electron mobility and on/off current ratio (Figure 8c and Figure 8d). In addition, we fabricated the multiple devices on planar and nanoscrolled 1L-MoS$_2$ and the transfer characteristics of these devices were provided in Figure S8 and Figure S9, respectively. Subsequently, the key device parameters are summarized in the Table 1. The threshold voltage (V$_t$) was calculated using Y-function method.[32] The mean V$_t$ values are 5.7 and 5.2 V for planar and nanoscrolled 1L-MoS$_2$, respectively. Earlier it was reported that V$_t$ was substantially decreased by increasing the number of MoS$_2$ layers.[33] However, the minute decrement in V$_t$ for nanoscrolled 1L-MoS$_2$ is attributed to weak interlayer coupling that exists among the constituent layers of the scroll. Generally, the on-state performance of the FET is influenced by the carrier mobility. We calculated the field-effect electron mobility (μ) from peak transconductance using equation (Eq. 1) [32]

$$\mu = \frac{L}{W \cdot C_{ox} \cdot V_{ds}} \left(\frac{\Delta I_{ds}}{\Delta V_{gs}}\right) \qquad (1)$$

where $C_{ox}$ is the gate insulator capacitance per unit area (11.51 × 10$^{-9}$ F·cm$^{-2}$ for SiO$_2$ of 300 nm), $L$ and $W$ are the length and width of the channel, respectively. The dimensions ($L \times W$) of the planar and scrolled device are 10 μm × 10 μm and 10 μm × 0.5 μm, respectively. The slope $\Delta I_{ds}/\Delta V_{gs}$ is calculated from the transfer characteristics curve in a linear regime at V$_{ds}$ of 1 V. The field-effect electron mobility is calculated to be ~10.5 and ~2400(±400) cm$^2$V$^{-1}$s$^{-1}$ for planar and nanoscrolled 1L-MoS$_2$, respectively. The on/off current ratios were 10$^6$ and 10$^7$ for planar and scrolled 1L-MoS$_2$, respectively. The above results indicate that nanoscrolled 1L-MoS$_2$ exhibits superior electronic properties compared to planar 1L-MoS$_2$. Moreover, a one-order higher current ratio makes the scrolled 1L-MoS$_2$ suitable for digital applications. The key device parameters of nanoscrolled 1L-MoS$_2$ FETs were compared with the existing literature (Table S5). The two-order higher on/off current ratio and one-order higher electron mobility in our nanoscrolled 1L-MoS$_2$ were observed compared to other reported nanoscrolled devices. As a matter of fact, the mobility value is the highest reported so far in any form of 1L-MoS$_2$, including that for top gated planar 1L-MoS$_2$, showing a mobility value of 1090 cm$^2$V$^{-1}$s$^{-1}$ (Table S6).[34] Thus, compact optoelectronic devices can be realized using nanoscrolled 1L-MoS$_2$, which retains the monolayer characteristics. Our results can be further extended to other 2D transition metal dichalcogenides monolayer structures to fabricate thermally stable scrolls.

Table 1: Summary of key device parameters of back-gated FET fabricated on planar and nanoscrolled 1L-MoS$_2$.

| Channel Material | I$_{on}$ (max) (μA/μm) | Threshold Voltage (V) | Mobility (cm$^{-2}$V$^{-1}$s$^{-1}$) | ON/OFF current ratio |
|---|---|---|---|---|
| 1L-MoS$_2$ film | 3 | 5.7 ± 0.2 | 10.5 ± 2.5 | 10$^6$ |
| Nanoscrolled 1L-MoS$_2$ | 45 | 5.2 ± 0.2 | 2400 ± 400 | 10$^7$ |

As discussed earlier, countless attempts are made to fabricate the scrolled MoS$_2$ with multiple layers which retain the monolayer nature.[8,9,11-13] Unfortunately, the fabricated scrolls behaved as multilayer MoS$_2$ where PL emission either disappeared or was suppressed. Moreover, the fabricated nanoscrolls have poor thermal stability where PL emission was suppressed after annealing.[11] In the present study, nano-scrolls showed enhancement of PL intensity and negligible shift compared to the planar 1L-MoS$_2$. In other words, the result signifies that nano-scrolls behave electronically in a similar manner to monolayers. The enhanced PL intensity is attributed to the non-interacting nature of constituent layers in the scroll. Interestingly, the monolayer behavior was further improved upon thermal annealing, which manifested their thermal stability. The properties of nanoscrolls are dependent on the arrangement of layers inside the scroll at their formation stage. The choice of solvent is most important as physical properties such as surface tension, evaporation rate, and dielectric constant strongly influence the scroll formation. The scrolls fabricated using the solvent with higher surface tension and evaporation rate increase the rolling speed. As a result, the stacking becomes imperfect and closely spaced layers are expected to form. Therefore, the interlayer interaction increases, and the scroll behaves as a multilayer. However, a solvent with optimum surface tension and evaporation rate can provide sufficient time for the formation of a scroll. In addition, the moderate dielectric constant of the solvent can further slowdown the formation of the scroll. In this context, the formation of a scroll using absolute ethanol is advantageous because of its optimum evaporation rate and moderate dielectric constant, which may slowdown the self-rolling process. As a result, one can expect a large separation between the constituent layers. The non-interacting nature of constituent layers might have originated from the large separation. Moreover, the constituent layers also retained the optical properties after a few months of fabrication (not shown here). It suggests that the scrolls have long-term stability in their optical properties. Therefore, our results find that absolute ethanol is a suitable solvent for the fabrication of nanoscrolls.

## 3. Conclusions

We strategically developed and demonstrated an effective and facile method to fabricate the thermally stable nanoscrolled 1L-$MoS_2$ from CVD-grown 1L-$MoS_2$ film. A suitable solvent with low surface tension and evaporation rate, ethanol, was used to fabricate the nanoscrolled 1L-$MoS_2$. The nanoscrolled 1L-$MoS_2$ retained the monolayer behaviour even though multilayer in morphology, which was confirmed by Raman spectroscopy. The nanoscrolled 1L-$MoS_2$ showed enhanced PL QY because of approximately two times increase in the absorbance and five times increase in the PL emission, compared to 1L-$MoS_2$. In addition, low-frequency Raman measurements confirm the negligible interlayer coupling among constituent layers in the insignificant presence of *S* and LB mode. Moreover, the KPFM study revealed insignificant variation in SP between monolayers and scrolled 1L-MoS$_2$, showing similar electronic properties of them. Furthermore, the enhancement of PL intensity (approximately six times) after annealing confirmed the thermal stability of fabricated nanoscrolls. The PL QY (2.95%) of nanoscroll 1L-MoS$_2$ was significantly higher than the reported the PL QY of untreated 1L-MoS$_2$. In addition, C-AFM measurements delineate that the nanoscrolled 1L-MoS$_2$ shows approximately six times higher photocurrent as compared to the 1L-MoS$_2$. FET device fabricated on nanoscroll 1L-MoS$_2$ has shown one order higher on/off ratio and two order higher mobility compared to the device fabricated on planar 1L-MoS$_2$. Therefore, fabricated nanoscrolls can be used for compact and high-performance optoelectronic devices.

## 4. Experimental section

The synthesis of 1L-MoS$_2$ flakes was carried out by the atmospheric pressure CVD technique. The MoS$_2$ phase was formed by using molybdenum trioxide (MoO$_3$) and S as precursor materials. The details of the synthesis steps can be found elsewhere.[2] The deposition of 1L-MoS$_2$ was carried out on SiO$_2$(300 nm)/Si susbstrate. The growth of planar and nanoscrolled 1L-MoS$_2$ was confirmed by optical microscopy. All morphological information, including the lateral dimension of the nanoscrolled 1L-MoS$_2$ was extracted from field emission scanning electron microscopy (FESEM; Supra 55, Zeiss, Germany) images. The thickness of 1L-MoS$_2$ and the height of the nanoscrolled 1L-MoS$_2$ were measured by the atomic force microscopy (AFM) technique (MultiView 4000, Nanonics Imaging Ltd, Israel). The AFM measurements were carried out in intermittent mode. Vibrational and emission properties were studied usingRaman and PL spectroscopy, respectively. Raman and PL measurements were performed by using a 532 nm laser line with a power of 0.01 mW.

Raman scattered light and PL emission light were dispersed through 2400 lines/mm and 1800 lines/mm grating, respectively. Subsequently, the dispersed light was collected by using a charged coupled device detector. Low-frequency Raman measurements were carried out by using a WiTec Alpha 300RA micro-Raman spectrometer. Here, the sample was excited with a 532 nm laser source, and scattered light was dispersed through 1800 lines/mm grating. A Bragg grating (Rayshield$^{TM}$) was employed to cut off the Rayleigh line.

Current mapping and I-V measurements were carried out using the conducting atomic force microscopy (C-AFM; Model No-NTEGRA II, NT-MDT, Russia). The whole metal Pt tip cantilever was used in contact mode. A typical load applied on the cantilever was ~ 50 nN during electrical measurement. Gold wire with pressure contact served as one electrode, and the AFM tip served as the second electrode. The sample was biased, and the AFM tip was grounded during the measurement. The current mapping was performed under the two-pass technique. In first-pass, the current was measured with the illumination of an in-built laser (650 nm) which was used for the feedback mechanism. The spill-over light from the feedback laser generated significant photo-carriers, which increased the net current during mapping. In the second pass, the feedback laser was off, and hence, the current mapping was performed in the dark. Similarly, I-V characteristics were also measured with and without the illumination of light from the feedback laser.

**PL QY measurements**

The PL QY of 1L-MoS$_2$ and nanoscrolled 1-MoS$_2$ was estimated using Rhodamine 6G (R6G) dye. The R6G shows 95% PL QY under 532 nm excitation. The steps involved in the estimation of PL QY can be found elsewhere.[4]

**Nanoscrolled 1L-MoS$_2$ based FET devices**

Both planar and nanoscrolled 1L-MoS$_2$-based back-gated FETs were fabricated with the same channel length of 10 μm. The FET fabrication protocols can be found elsewhere.[28] The electrodes (Au (200 nm)/Ti (100 nm)) were prepared via the lift-off method by direct laser writer (M/s Microtech, Italy, Model No: LW405B) and wet chemical methods. The electrodes were deposited by dual-target DC magnetron sputtering. The FET characteristics were measured by source measurement unit (Keysight, Model-B2902A).

**Acknowledgment**

We thank Dr. Suman Paul, attocube Systems AG, Eglfinger Weg 2, 85540 Haar, for nano FTIR measurements. We also thank the Director, MSG, and the Director, IGCAR, for their constant encouragement.

**Conflict of Interest**

The authors declare no conflict of interest.

**Figures**

**Figure 1**

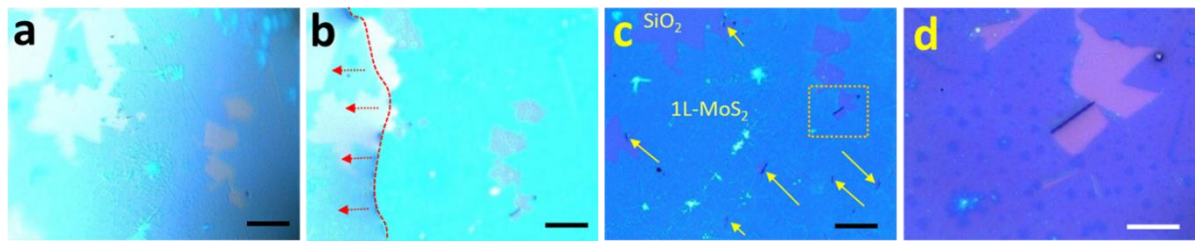

**Figure 1.** Step-by-step monitoring the fabrication of nanoscrolled 1L-MoS$_2$. (a) The surface of 1L-MoS$_2$ under the drop of absolute ethanol. (b) Shrinkage of ethanol drops during its evaporation (boundary marked with red dotted curve and arrow marks indicate the direction of shrinkage). (c) Multiple scrolls formed after ethanol evaporation (pointed by arrow marks). The given scroll selected for further studies was highlighted with a dotted box. Scale bar: 20 μm (d) High magnification image of one of the nanoscrolls. Scale bar: 10 μm.

**Figure 2**

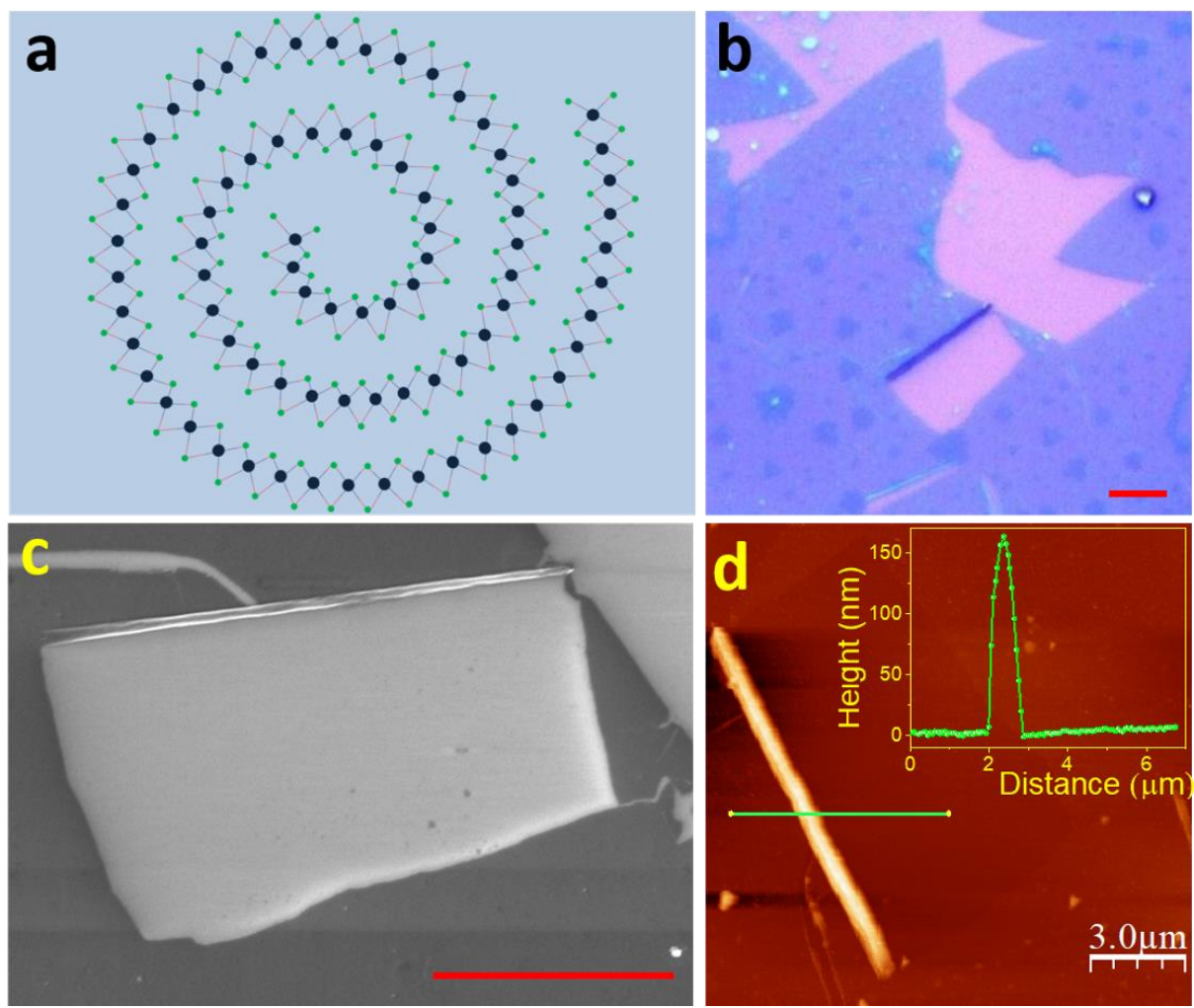

**Figure 2.** a) Schematic representing a cross-sectional view of 1L-MoS$_2$ nanoscroll. The optical image (b) and its respective FESEM image (c) of the particular scroll. Scale bar: 5 μm. d) The topographic image of the same nanoscroll. Inset shows the height profile across the scroll. The line profile was taken along the green line shown (d).

**Figure 3**

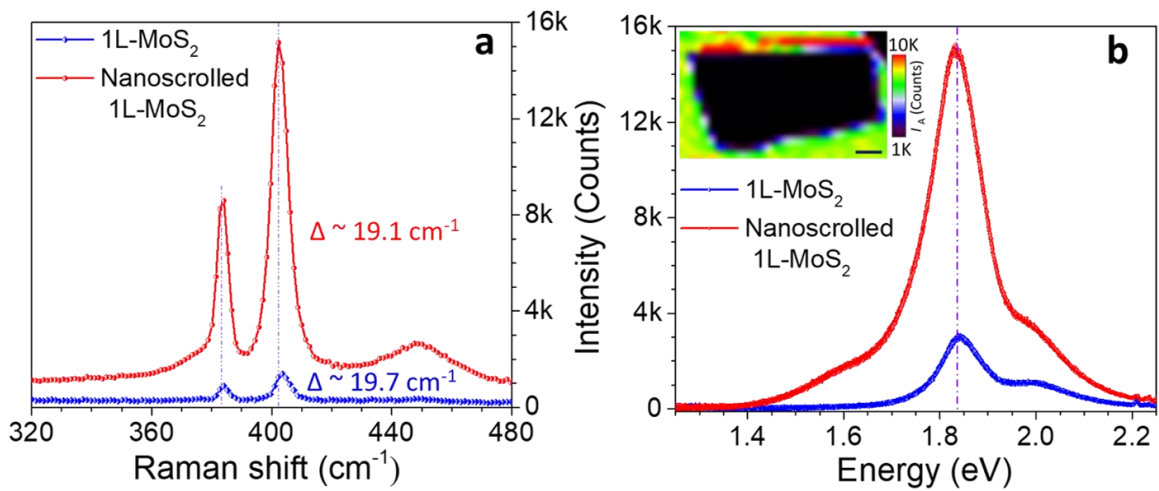

**Figure 3.** Comparative optical characterization of planar and nanoscrolled 1L-MoS$_2$. (a) Raman and (b) PL spectra of planer and scrolled MoS$_2$. The Δ value in both planar and nanoscrolled MoS$_2$ confirms the monolayer nature of MoS$_2$. Dotted lines are guide to the eye. Insets of (b) show the intensity map at a peak energy of A-exciton (1.84 eV). Scale bar (inset figure): 2 μm.

**Figure 4**

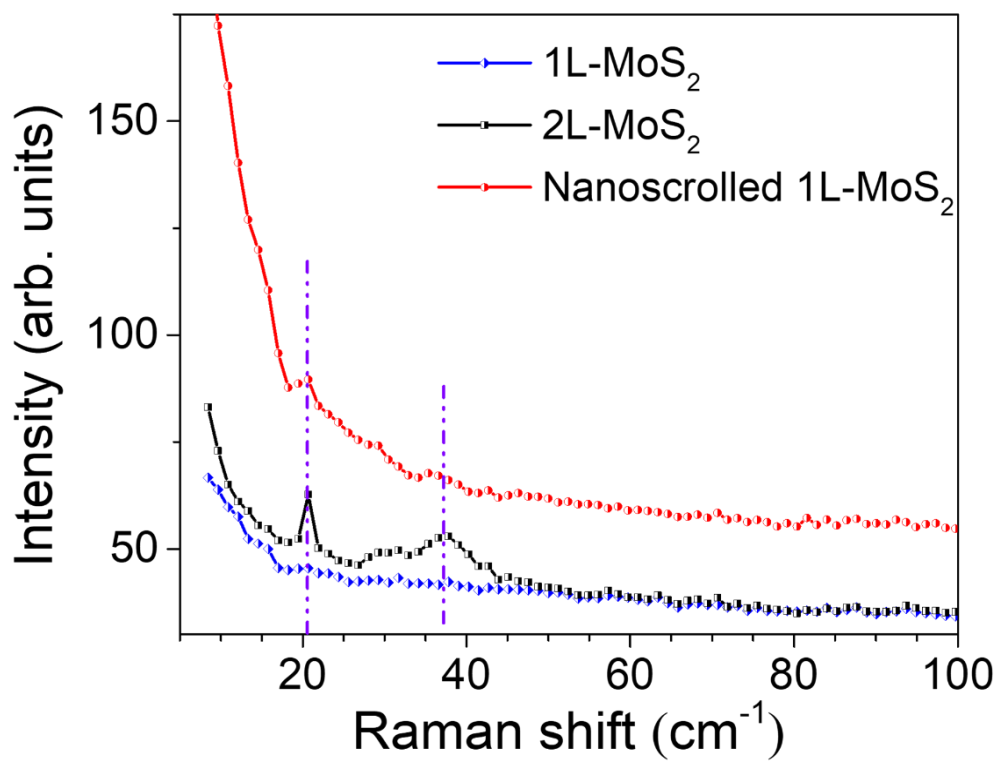

**Figure 4.** Low-frequency Raman spectra of 1L-, 2L-MoS$_2$ and nanoscrolled 1L-MoS$_2$.

**Figure 5**

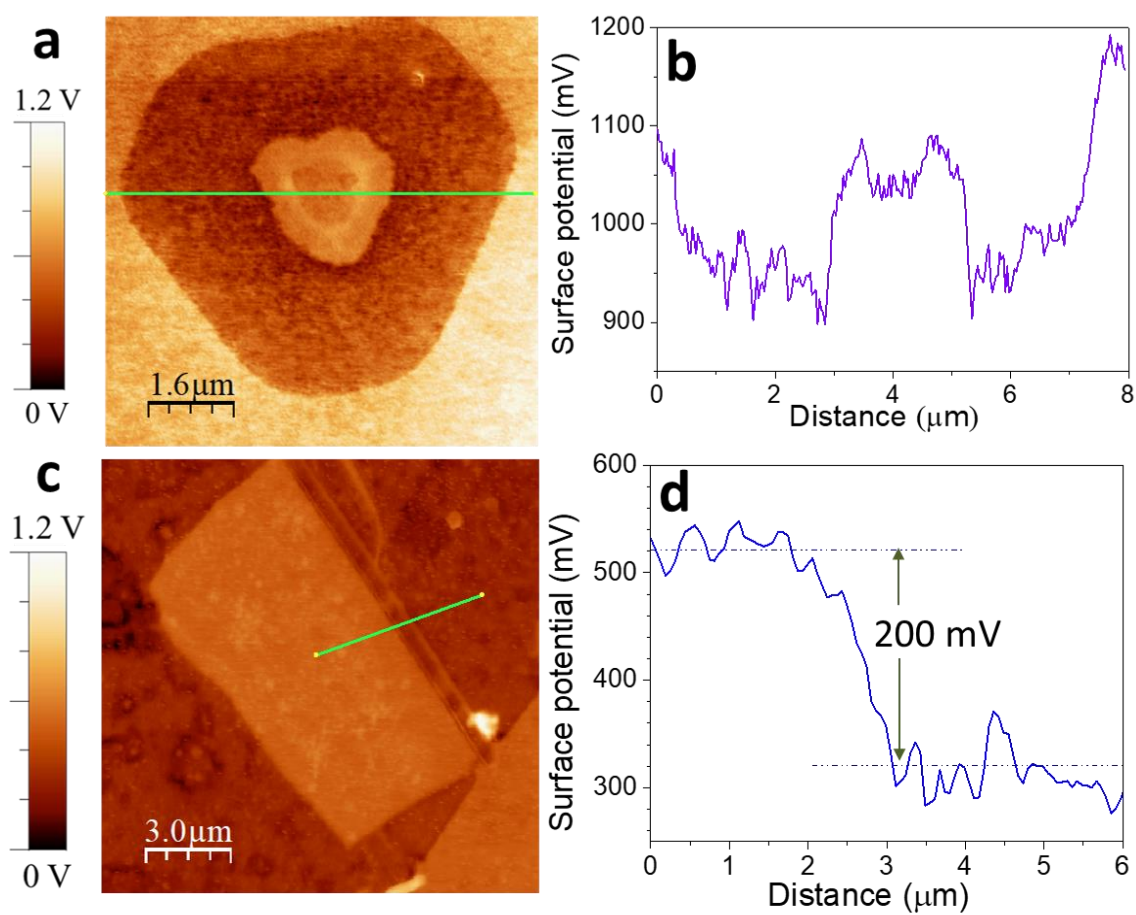

**Figure 5.** Surface potential analysis of 1L-, 2L- and nanoscrolled 1L-MoS$_2$ using KPFM. a) KPFM image showing the spatial variation of SP in a flake containing, mono-, bi- and multi-layered MoS$_2$. b) Line profile across flake. The line profile was taken along the green line (a). c) SP map of nanoscrolled 1L-MoS$_2$ region. d) Line profile of SP provided was taken along the green line in (c).

**Figure 6**

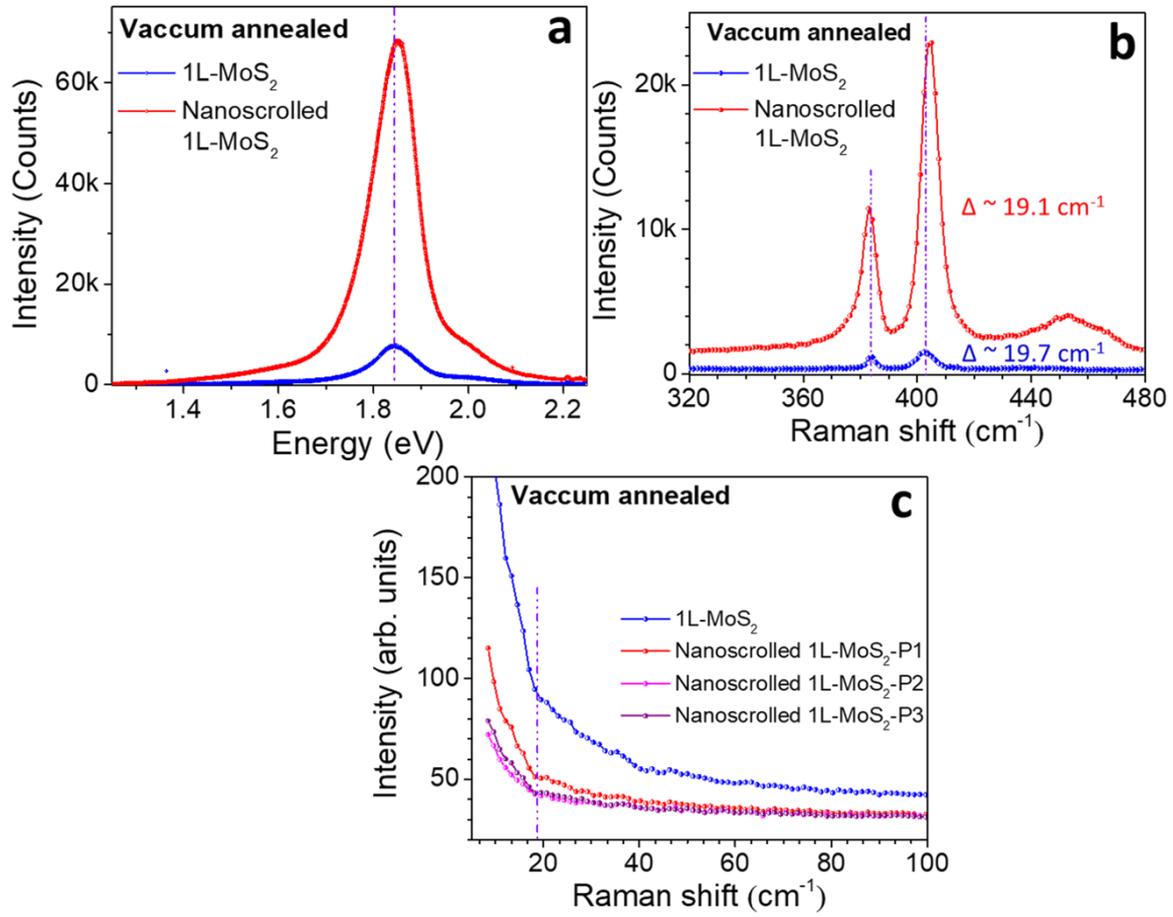

**Figure 6.** Raman and PL analysis of vacuum annealed nanoscrolled 1L-MoS$_2$. Photoluminescence (a), Raman (b) and LF Raman (c) spectra of vacuum annealed nanoscrolled 1L-MoS$_2$. LF Raman measurements were carried out at three arbitrary positions (P1, P2, and P3) on nanoscrolled 1L-MoS$_2$.

**Figure 7**

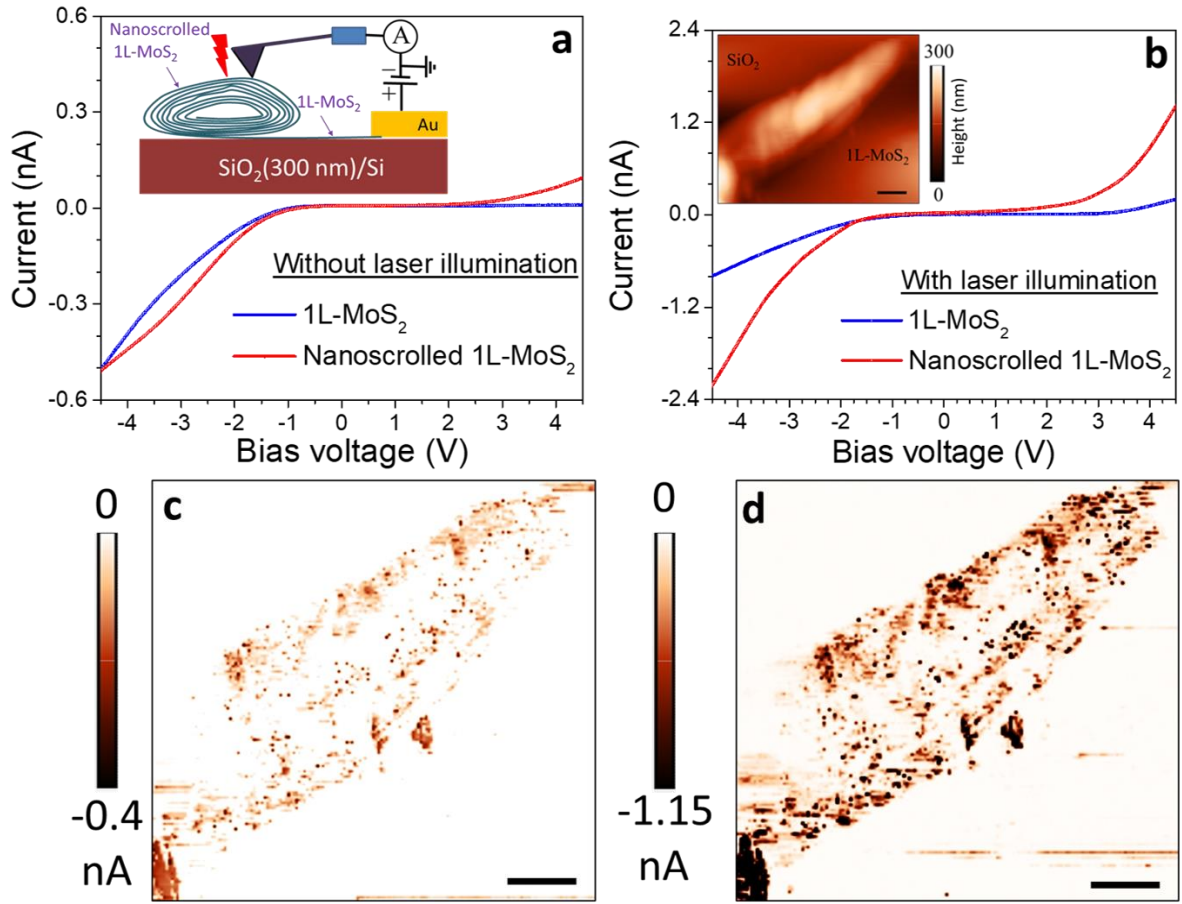

**Figure 7.** Conductive AFM-based I-V characteristics of 1L-MoS$_2$ and scrolled 1L-MoS$_2$ without (a) and (b) with spill-over light from feedback laser. Inset of (a) represents the schematic of C-AFM measurement. The Inset of (b) corresponds to the topography of the scroll. The C-AFM image of scrolled 1L-MoS$_2$ region under (c) dark and (d) illumination. Scale bar: 1 μm.

**Figure 8**

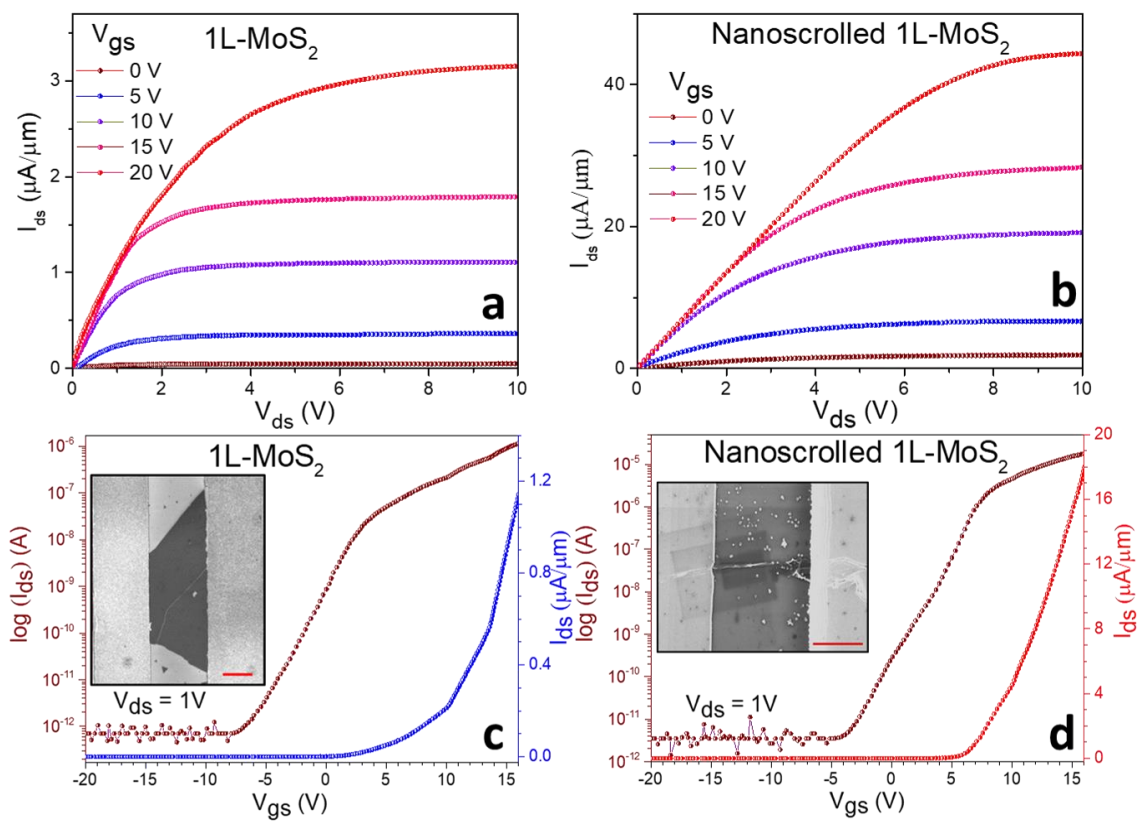

**Figure 8.** Transfer and output characteristics of FET devices fabricated on planar and nanoscrolled 1L-MoS$_2$. Output characteristics of (a) planar 1L-MoS$_2$ FET and (b) nanoscrolled 1L-MoS$_2$ FET. Transfer characteristics of (c) planar 1L-MoS$_2$ FET and (d) nanoscrolled 1L-MoS$_2$ FET. The insets of (c) and (d) show the FESEM image of typical FET devices fabricated on planar and scrolled 1L-MoS$_2$, respectively. Scale bar: 5 μm.